\documentclass[11pt]{article}
\usepackage{amsmath,amssymb,amsthm,amscd,tabularx,array,fancyhdr}
\newfont{\extra}{msbm10 scaled\magstep1}

\newcommand{\extr}[1]{\mbox{\extra #1}}

\newcommand{\sect}[1]{\setcounter{equation}{0}\section{#1}}

\def\be{\begin{equation}}
\def\ee{\end{equation}}
\def\bea{\begin{eqnarray}}
\def\eea{\end{eqnarray}}

\def\R{\extr R}

\newcommand{\bicross}{\triangleright\!\!\!\blacktriangleleft}

\newcommand{\RL}{\triangleright\!\!\!\blacktriangleleft}

\newcommand{\LEMO}{>\!\!\!\triangleleft}

\newcommand{\ract}{\triangleleft}
\newcommand{\lcact}{\blacktriangleleft}

\begin{document}

\begin{center}{ \LARGE \bf
Dynamical Systems  \\[0.4cm]
and \\[0.4cm]
Quantum Bicrossproduct Algebras}
\end{center}
\vskip0.25cm

\begin{center}
Oscar Arratia $^1$ and Mariano A. del Olmo $^2$
\vskip0.25cm

{ \it $^{1}$ Departamento de  Matem\'atica Aplicada a la  Ingenier\'{\i}a,  \\
\vskip0.10cm
$^{2}$ Departamento de  F\'{\i}sica Te\'orica,\\
\vskip0.10cm
 Universidad de  Valladolid, 
 E-47011, Valladolid,  Spain}

\vskip0.15cm

E. mail: oscarr@wmatem.eis.uva.es, olmo@fta.uva.es
 
\end{center}
 
\vskip1.5cm
\centerline{\today}
\vskip1.5cm

\begin{abstract}
We present a unified study of some aspects of  quantum bicrossproduct algebras of 
inhomogeneous Lie algebras, like Poincar\'e, Galilei and Euclidean in $N$ dimensions. The action
associated to the bicrossproduct structure allows to obtain a nonlinear action over a new
group linked to the translations.  This new nonlinear action associates a dynamical
system to each generator which is the object of study in this paper.
\end{abstract}
\newpage

\sect{Introduction}

In a series of papers \cite{olmo98}--\cite{olmo01a} we have dealt with the problem of the
construction of induced representations of quantum inhomogeneous algebras. In particular, in
Ref.~\cite{olmo01} we have focussed our attention on quantum Hopf algebras having the structure
of bicrossproduct ${\cal H}= U({\cal K}) \bicross U_z({\cal L})$, with $U({\cal K})$ a
cocommutative Hopf algebra, $U_z({\cal L})$ a commutative but noncocommutative Hopf algebra and
$\cal K$ and $\cal L$ Lie algebras \cite{majid}. Remember that this bicrossproduct structure
is the deformed counterpart of the semidirect product of Lie groups ($H=L\odot K$). 
In this paper we want to profit of some of the techniques
developed in the above mentioned papers to obtain relevant information about some aspects related
with the bicrossproduct algebras object of our study.

We shall reinterpret the above  bicrossproduct structure as  
$H= U({\cal K}) \bicross {\rm Fun}({L_z})$, because the commutativity of
${\cal L}$ allows to identify it with the algebra of functions over a certain group $L_z$. The
bicrossproduct structure determines an action of $U({\cal K})$ on  ${\rm Fun}({L_z})$ which at
the level of groups originates a nonlinear action of the group $K$ on $L_z$. At the
infinitesimal level this last action is described by vector fields associated to the generators
of ${\cal K}$. These vector fields give rise to  oneparameter flows some of them linear
(`nondeformed') and other nonlinear  (`deformed'). In other words, we can study
some dynamical systems associated to this action. 

As it is well known in the nondeformed case, the homogeneous space $X=H/K$ is diffeomorphic to
$\R^N$ which is associated to $L$, being $N$ the dimension of $L$. In the quantum case
the homogeneous space is now identified with ${\rm Fun}({L_z})$. However, we can study the
nonlinear action of $K$ over $L_z$. Note that in the limit $z\to 0$ we recover the linear
action of $K$ on $X$.

We will consider the family of inhomogeneous algebras related by graded contractions with the
compact algebra $so(N+1)$ \cite{olmo93,olmo94}. They are called inhomogeneous Cayley--Klein
algebras. Among the elements of this family we find the Poincar\'e and the Galilei algebras in
$(N-1,1)$ dimensions and the Euclidean algebra in $N$ dimensions.  
The bicrossproduct structure that share these quantum algebras \cite{olmo97} allows to present a
unified study of the properties mentioned above.

The organization of the paper is as follows. Section \ref{standardeformationn} presents a brief
review about the inhomogeneous Cayley--Klein algebras, their quantum deformations and their
bicrossproduct structure. Next section, the most important of the work, is devoted to  compute
the flow associated to the action, the invariant under this action that coincides with the
Casimir, and the dynamical systems associated. 
 In section
\ref{deformacionestandarso2} we present, as an example, the case of
$N=3$ to illustrate the ideas introduced in the previous section. 
We finish with some conclusions and remarks.

\sect{Quantum Cayley-Klein algebras  
$U_z(\mathfrak{iso}_{\omega_2, \omega_3, \ldots,\omega_N}(N))$}\label{standardeformationn}
 
The  family of  Cayley--Klein  pseudo-orthogonal algebras 
is a set of ${(N+1)N}/{2}$ dimensional real Lie  algebras 
characterized by $N$ real parameters $(\omega_1,\omega_2, \ldots, \omega_N)$ and 
 denoted $\mathfrak{so}_{\omega_1,\omega_2, \ldots, \omega_N}(N+1)$ \cite{olmo93,olmo94}.
In an appropriate basis $(J_{ij})_{0\leq i < j \leq N}$ the nonvanishing commutators are
$$
[J_{ij}, J_{ik}]= \omega_{ij}J_{jk} , \qquad
   [J_{ij}, J_{jk}]= - J_{ik} ,\qquad
   [J_{ik}, J_{jk}]= \omega_{jk}J_{ij},
$$
with  the subindices verifying 
$0<i<j<k<N$ and $\omega_{ij}={\displaystyle \prod_{s=i+1}^j \omega_s}$.
The generators can be rescaled in such a
way that the parameters $\omega_i$ only take  the values $1,0$ and $-1$.
When all  the  $\omega_i$'s are different to zero the algebra
$\mathfrak{so}_{\omega_1,\omega_2, \ldots, \omega_N}(N+1)$
is isomorphic to  some of the  pseudo-orthogonal algebras
$\mathfrak{so}(p,q)$ with $p+q=N+1$ and $p\geq q >0$. If some of the coefficients
$\omega_i$ vanishes the corresponding  algebra is inhomogeneous and can be obtained
from $\mathfrak{so}(p,q)$ by means  of  a sequence of contractions. 
In the particular case of $\omega_1=0$, the algebras
$\mathfrak{so}_{0,\omega_2, \ldots, \omega_N}(N+1)$ can be realized as algebras of  groups
of  affine transformations on $\mathbb{R}^N$ \cite{olmo93}.  In this case, the generators
$J_{0i}$ are denoted by $P_i$ stressing, in this way, its role as 
generators of  translations. The   remaining generators $J_{ij}$ originate
compact and `noncompact' rotations.
These inhomogeneous algebras, henceforth denoted by
$\mathfrak{iso}_{\omega_2, \ldots, \omega_N}(N)$, are characterized by
the following nonvanishing commutators
$$\begin{array}{llll}
  &[J_{ij}, J_{ik}]= \omega_{ij}J_{jk},  \qquad
      &[J_{ij}, J_{jk}]= -J_{ik}, \qquad
      &[J_{ik}, J_{jk}]= \omega_{jk} J_{ij},\\[0.2cm]
 &[J_{ij}, P_i] = P_j, \qquad &[J_{ij}, P_i]= - \omega_{ij} P_i, \qquad &\qquad\qquad 
1\leq i<j<k\leq N .
\end{array}$$

In \cite{olmo93a,olmo94a}  simultaneous standard deformations (i.e. their associated classical
$r$--matrices are quasi-triangular \cite{ChP}) for all the  enveloping algebras 
$U(\mathfrak{so}_{\omega_1,
\omega_2}(3))$ and 
$U(\mathfrak{so}_{\omega_1, \omega_2, \omega_3}(4))$, respectively, were introduced. 
In \cite{olmo94b} the case of 
$U(\mathfrak{iso}_{\omega_2, \omega_3, \omega_4}(4))$ was considered, and in 
 \cite{olmo95} the general case 
$U(\mathfrak{iso}_{\omega_2, \omega_3,\ldots,\omega_N}(N))$ (all of them standard
deformations).

It was proved  in \cite{olmo97}  that the standard quantum Hopf algebras
$U_z(\mathfrak{iso}_{\omega_2, \omega_3,\ldots, \omega_N }(N))$
have a structure of bicrossproduct.
Using a basis adapted to the bicrossproduct structure we can describe together all 
these quantum algebras 
$U_z(\mathfrak{iso}_{\omega_2, \omega_3,\ldots, \omega_N }(N))$. In order to avoid
repetitions we use the following convention: the   variation rank of  
$i,j,k$ is $1,\ldots, N-1$ and the  index
$N$ is treated separately.  Besides, when two  indices
$i,j$ appear in a generator it is  assumed that $i<j$.
The commutation relations are
 \begin{equation} \label{conmuso..}
 \begin{split}
    &  [P_i,P_j]=0, \quad\qquad \qquad [P_i, P_N]=0, \\[0.2cm]
     & [J_{ij}, J_{ik}]= \omega_{ij} J_{jk}, \qquad \ 
       [J_{ij}, J_{jk}]= - J_{ik}, \qquad\qquad
       [J_{ik}, J_{jk}]=  \omega_{jk} J_{ij},\\[0.2cm]
     & [J_{ij}, J_{iN}]= \omega_{ij} J_{jN}, \qquad
       [J_{ij}, J_{jN}]= - J_{iN}, \qquad\quad\
       [J_{ik}, J_{jN}]=  \omega_{jN} J_{ij},\\[0.2cm]
     & [J_{ij}, P_k]= \delta_{ik} P_k- \delta_{jk} \omega_{ij} P_i,  \qquad\qquad \qquad \qquad
       \quad \ \  [J_{ij}, P_N]= 0, \\[0.3cm]
     &  [J_{iN}, P_j]= \delta_{ij} \left( \frac{1-e^{-2z P_N}}{2z}
           - \frac{z}{2} \sum_{s=1}^{N-1} \omega_{sN} P_s^2\right) +
       z \omega_{iN}P_i P_j, \qquad [J_{iN}, P_N]= -\omega_{iN} P_i\ ;         
   \end{split}\end{equation} 
and the   coproduct is given by 
\begin{equation} \begin{split}
    &  \Delta(P_i)= P_i \otimes 1+ e^{-z P_N} \otimes P_i, \qquad
     \Delta(P_N)= P_N \otimes 1+ 1 \otimes P_N, \\[0.3cm]
    &  \Delta(J_{ij})=  J_{ij}  \otimes 1 + 1 \otimes J_{ij}, \\[0.3cm]
    &  \Delta(J_{iN})=  J_{iN}   \otimes 1 + e^{-z P_N} \otimes J_{ij}
           + z \sum_{s=1}^{i-1} \omega_{iN} P_s \otimes J_{si}
           - z \sum_{s=i+1}^{N-1} \omega_{sN} P_s \otimes J_{is}\ . \nonumber
    \end{split} \end{equation} 
The  bicrossproduct structure
$U_z(\mathfrak{iso}_{\omega_2, \omega_3,\ldots, \omega_N }(N))= {\cal K} \RL {\cal L}$,
with ${\cal K}= U(\mathfrak{so}_{\omega_2, \omega_3,\ldots, \omega_N }(N))$ and $\cal L$ 
the  commutative Hopf subalgebra generated by $P_1, P_2, \ldots, P_N$, is described by
the right  action of $\cal K$ over  $\cal L$
$$
  P_i \ract J_{jk}= [P_i, J_{jk}], \qquad j<k, \; i,j,k=1,2,\ldots,N,
$$
with the commutators given by (\ref{conmuso..}), and  the  left coaction of  $\cal L$
over  $\cal K$, whose expression   over  the generators of   $\cal K$ is
\begin{equation}
\begin{split}
     & J_{ij} \lcact= 1 \otimes J_{ij}, \\
     & J_{iN} \lcact= e^{-z P_N} \otimes J_{iN}
           + z \sum_{s=1}^{i-1} \omega_{iN} P_s \otimes J_{si}
           - z \sum_{s=i+1}^{N-1} \omega_{sN} P_s \otimes J_{is}.\nonumber
     \end{split}\end{equation}

\sect{Oneparameter flows}\label{oneparametricflows}

In \cite{olmo97} the  algebra $U_{z}(T_N)$ was 
considered as a  noncommutative deformation of the Lie algebra of the group of 
translations of  $\mathbb{R}^N$. However, here we can profit  the commutativity of  $U_{z}(T_N)$
for  interpreting it  as   the  algebra of  functions over a group, in such a way that we have
the following  bicrossproduct decomposition 
$$
U_z(\mathfrak{iso}_{\omega_2, \omega_3,\ldots, \omega_N }(N))=
U(\mathfrak{so}_{\omega_2, \omega_3,\ldots, \omega_N }(N)) \RL F(T_{z,N}),
$$
where  $T_{z, N}$ is the   space
$\mathbb{R}^N$ equipped with   the  composition law 
 \begin{equation}\begin{split}
   & (\alpha'_1,\alpha'_2, \ldots,\alpha'_{N-1},\alpha'_N )
    (\alpha_1,\alpha_2, \ldots,\alpha_{N-1},\alpha_N )= \\[0.2cm]
   & \hspace{2cm}
   (\alpha'_1 + e^{-z \alpha'_N} \alpha_1,
   \alpha'_ 2+ e^{-z \alpha'_N} \alpha_2, \ldots,
   \alpha'_{N-1}+e^{-z \alpha'_N} \alpha_{N-1},\alpha'_N +\alpha_N), \nonumber
    \end{split}\end{equation}
that equips it with  a structure of $N$--dimensional Lie group. 
The  group  $T_{z, N}$ has the  structure  of   semidirect product of the
additive groups $\mathbb{R}^{N-1}$ and $\mathbb{R}$
$$
    T_{z, N} \equiv \mathbb{R}^{N-1} \LEMO \, \mathbb{R},
    \qquad (a',b')(a,b)= (a'+ a \ract {b'}^{-1}, b'+b),
$$
where the right action  of  $\mathbb{R}$ over $\mathbb{R}^{N-1}$
is given by means of the usual  product  over  each component,
$$
      a \ract b= e^{z b} a,
     \qquad a \in \mathbb{R}^{N-1}, \quad b \in \mathbb{R}.
$$
The generators $P_i$ of $U_{z}(T_N)$ give in this context  a global chart over 
$T_{z, N}$,
$$
        P_i(\alpha)= \alpha_i, \qquad \alpha \in T_{z, N}.
$$
 The  structure of  
$U (\mathfrak{so}_{\omega_2, \omega_3,\ldots, \omega_N }(N))$--module algebra of 
$F(T_{z, N})$ implies that an  action of the 
group $SO_{\omega_2, \omega_3,\ldots, \omega_N }(N)$ on  $T_{z, N}$  is defined. At
the  infinitesimal level this  action  is  described  by the  vector fields
 \begin{equation} \label{fieldsison}
     \begin{split}
    \hat{J}_{ij}= & -P_j \frac{\partial}{\partial P_i} +
                  \omega_{ij} P_i \frac{\partial}{\partial P_j},  \\[0.3cm]
    \hat{J}_{iN}= &  \sum_{j=1}^{N-1} - \left[
    \delta_{ij} \left(\frac{1- e^{-2z P_N}}{2z}-
     \frac{z}{2} \sum_{s=1}^{N-1} \omega_{sN} P_s^2 \right)+
      z \omega_{iN} P_i P_j \right]  \frac{\partial}{\partial P_j}
         + \omega_{iN} P_i    \frac{\partial}{\partial P_N} \\[0.2cm]
        =& \sum_{j\not=i,N} - z \omega_{iN} P_i P_j
           \frac{\partial}{\partial P_j}-
            \left[ \frac{1- e^{-2z P_N}}{2z}-
               \frac{z}{2} \sum_{s=1}^{N-1} \omega_{sN} P_s^2 
+  z \omega_{iN} P_i^2 \right] \frac{\partial}{\partial P_i} 
    + \omega_{iN} P_i    \frac{\partial}{\partial P_N}.
\end{split} \end{equation}

Since only the  generators $J_{1N},J_{2N},\dots, J_{N-1\;N}$ have deformed  action 
the  integration of  the equations of  the fields 
$\hat{J}_{ij}$ is immediate  and gives the well known linear flows
 \begin{equation} \label{flujnodefison}
   \Phi_{ij}^t(\alpha_1, \ldots,  \alpha_i,\ldots,  \alpha_j, \ldots,  \alpha_N )=
  (\alpha_1, \ldots, \alpha_{i-1},  \alpha'_i, \alpha_{i+1} \ldots,
    \alpha_{j-1},  \alpha'_j, \alpha_{j+1} \ldots,  \alpha_N ),
 \end{equation}
with
$$
\alpha'_i= C_{\omega_{ij}}(t) \alpha_i - S_{\omega_{ij}}(t) \alpha_j,\qquad
   \alpha'_j= \omega_{ij} S_{\omega_{ij}}(t) \alpha_i + C_{\omega_{ij}}(t) \alpha_j,
$$
where 
$$
  C_{\omega}(t)=\frac{e^{\sqrt{-\omega} t} + e^{-\sqrt{-\omega} t}}{2},
 \qquad
  S_{\omega}(t)= \frac{e^{\sqrt{-\omega} t} - e^{-\sqrt{-\omega} t}}{2 \sqrt{-\omega}}.
$$
So,  we have simple compact or noncompact rotations in the $ij$  plane.  

The  computation of the flows associated to  the `deformed' fields  $\hat{J}_{iN}$
requires a more careful analysis.
Let us start by obtaining their invariants.
Supposing that the differential form
\begin{equation}\label{differentialform}
\eta= \sum_{s=1}^N \mu_s \; dP_s,
\end{equation}
verifies $\hat{J}_{iN} \rfloor \eta=0$,
 the  following equation is obtained
\begin{equation} \label{ecu1forma}
\sum_{j\not=i,N} z \omega_{iN} P_i P_j \mu_j +
       \left[ \frac{1- e^{-2z P_N}}{2z}-
          \frac{z}{2} \sum_{s=1}^{N-1} \omega_{sN} P_s^2 
+ z \omega_{iN} P_i^2 \right] \mu_i - \omega_{iN} P_i    \mu_{N}=0.
\end{equation}
Using this expression   $(N-1)$ invariant functions are obtained as follows. For the first
invariant we choose 
$$
    \mu_s= \omega_{sN} P_s \tau, \qquad s=1,2,\ldots,N-1,
$$
with $\tau$ a function to be evaluated. Hence,  equation (\ref{ecu1forma}) 
reduces to
\begin{equation}\label{ecu1forma1}
 \omega_{iN}P_i\left[ \frac{1- e^{-2z P_N}}{2z}+
   \frac{z}{2} \sum_{s=1}^{N-1} \omega_{sN} P_s^2
     \right] \tau - \omega_{iN}P_i \mu_N=0 .
\end{equation} 
From equation (\ref{ecu1forma1}) we find the value of $\mu_N$  obtaining  the differential
form
$$
 \eta= \tau \left[\sum_{j=1}^{N-1} \omega_{jN} P_j \; dP_j 
+ \left( \frac{1- e^{-2z P_N}}{2z}+
     \frac{z}{2} \sum_{s=1}^{N-1} \omega_{sN} P_s^2 \right) dP_N \right] ,
$$
where  $\tau$ plays the role of integration factor.  Solving the case $N=2$ we get
$\tau = 2 e^{z P_N}$,
that it is proved to be valid for every $N$. 
The  integration of  the equations
$$
\frac{\partial h}{\partial P_s}= \mu_s, \qquad 1\leq s\leq N ,
$$
gives  $\eta=dh$. 
By an  appropriate choice of the integration  constant, in order to have a well behaviour
in the   limit $z \rightarrow 0$, we obtain
\begin{equation} \label{finvcasimir}
 h_{\omega,z}=  \sum_{j=1}^{N-1}  \omega_{jN} P_j^2 e^{z P_N}
+ \frac{\cosh(z P_N)-1}{\frac{z^2}{2}}.
\end{equation}
This function is, in fact, invariant under the  action of  all the generators $J_{ij}$.
Indeed, it belongs to the center of the algebra
$U_z(\mathfrak{iso}_{\omega_2, \omega_3, \ldots, \omega_N}(N))$ and  is the  
Casimir $C_z$ given in \cite{olmo97},  but now it appears in a natural way.
   
To obtain the other $N-2$ invariants we start 
fixing  $k \in \{1,2, \ldots, N-1 \}-\{i\}$ and taking $\mu_j=0$ if
$j\not= k, N$, we get from expression (\ref{differentialform})  the  differential form
$  \eta_k= \mu_k dP_k + \mu_N dP_N$.
Condition (\ref{ecu1forma}) applied to $\eta_k$
establishes a relationship between $\mu_k$ and $\mu_N$ that allows to write
$$
   \eta_k= \mu_k dP_k + \mu_k z P_k dP_N.
$$
Choosing  $\mu_k= e^{z P_N}$  the  differential form is exact,
that is,  $\eta_k= d h^{iN,k}_{\omega, z}$, with
\begin{equation}\label{invariantesh}
  h^{iN,k}_{\omega, z}= P_k e^{z P_N},
   \qquad k\in \{1,2,\ldots, N-1 \}- \{i\}.
\end{equation}
To obtain the integral curves of  $\hat{J}_{iN}$ it is necessary to  solve
the   system of  $N$ differential equations
\begin{equation}
\begin{split}
     & \dot{\alpha}_j= - z \omega_{iN}\alpha_i \alpha_j, 
\qquad \qquad j \not= i, N, \\[0.2cm]
     & \dot{\alpha}_i=  - \frac{1- e^{-2z \alpha_n}}{2z}+
        \frac{z}{2} \sum_{s=1}^{N-1} \omega_{sN} \alpha_s^2 -
          z \omega_{iN} \alpha_i^2, \\[0.2cm]
     & \dot{\alpha}_N=    \omega_{iN} \alpha_i.\nonumber
 \end{split}
\end{equation} 
The invariants $h^{iN,k}_{\omega, z}$ (\ref{invariantesh}) allow to
remove $N-2$ degrees of  freedom, from
$h^{iN,k}_{\omega, z}(\alpha)=\alpha_k e^{z \alpha_N}=\beta_k$ we obtain
$\alpha_k=\beta_k e^{- z \alpha_N}$, 
restricting the study of the $N$--dimensional system
to the following family of  2--dimensional systems depending on  the $N$
parameters $\beta_k$,  $\omega$ and $z$
\begin{equation} \label{sisred}
 \begin{split}
  & \dot{\alpha}_i= -\left[  \frac{1- e^{-2z \alpha_N}}{2z}-
  \frac{z}{2} \left(
     \sum_{s\not= i, N} \omega_{sN} \beta_s^2\right) e^{-2z \alpha_N} +
        \frac{z}{2} \omega_{iN} \alpha_i^2 \right], \\
     & \dot{\alpha}_N=    \omega_{iN} \alpha_i.
\end{split}\end{equation}
 Due to the  way  in which  the parameters $\beta_k$ appear grouped, 
 the  set of  systems (\ref{sisred}) only depends on three parameters
$z$, $\omega_{iN}$ and $\rho={ \sum_{s\not= i, N}} \omega_{sN} \beta_s^2$.
The  function $h_{\omega,z}$ (\ref{finvcasimir})  gives the following
invariant for the   system (\ref{sisred})
\begin{equation}  \label{finvcasimirbid}
      \omega_{iN} \alpha_i^2 e^{z \alpha_N} +
         \rho e^{-z \alpha_N} +
        \frac{\cosh(z \alpha_N)-1}{\frac{z^2}{2}}. 
\end{equation}
The  description of  the systems when $z=0$ is trivial, since it reduces to the
study of linear systems analogue to those of the fields
$\hat{J}_{ij}$. If $z$ does not vanish the equations
may be rescaled considering
$$
x(t)=  z  \alpha_i(t),\qquad
  y(t)=   z \alpha_N(t),
$$
and  setting $a= \omega_{iN}$,
$b-1=z^2 \rho= z^2 {\displaystyle \sum_{s\not= i, N}} \omega_{sN} \beta_s^2$
the   system becomes
  \begin{equation}    \label{ecutrayiso}
       \dot{x}=  -\frac{1}{2} a x^2 -\frac{1}{2} +\frac{1}{2} b e^{-2y} ,\qquad
       \dot{y}=  \, a x .
\end{equation}
In this form the limit $z \to 0$ cannot be studied, but in 
advantage it depends on only two parameters. The  possibility of  reabsortion of 
the parameter $z$ is followed from the fact that all the Hopf algebras
$U_z(\mathfrak{iso}_{\omega_2,\omega_3, \ldots,
\omega_N}(N))$ are isomorphic (for fixed values  of  the parameters $\omega_s$) 
whenever $z$ is nonzero.
The  function (\ref{finvcasimirbid})
gives rise to the following invariant of  (\ref{ecutrayiso})
 \begin{equation} \label{invaab}
        h_{a,b}= a x^2  e^{y}+  e^{y} +  b e^{-y}.
\end{equation}
The  research of fixed points of the system reveals that:
 \begin{itemize}
   
\item if $b\leq 0$ the system has not  equilibrium points;
   
\item if $b > 0$ there are three possibilities:
          
\begin{itemize}
            
\item if $a<0$ then there is only one fixed point  $(0,\frac{1}{2} \ln b)$  of hyperbolic
character,
            
\item if $a=0$ then all the points like $(x,\frac{1}{2} \ln b)$
are fixed points,
            
\item if $a>0$ there is only one equilibrium point
$(0,\frac{1}{2} \ln b)$  of  elliptic character.
\end{itemize}
\end{itemize}
Let us go to analyze in detail the case $a>0$ and
$b>0$. Here the invariant (\ref{invaab}) has a global minimum of value
$2 \sqrt{b}$ at 
$(0, \frac{1}{2}\ln b)$  and it is easy to check that $h_{a,b}$ takes
arbitrarily high values over  points going to   infinity in any direction. Since the
orbits of the system (\ref{ecutrayiso}) are the level curves of  $h_{a,b}$
all the  orbits are bounded.
 Note that the   point   of  equilibrium disappears in the  
limit $b \rightarrow 0$.
Let us consider  the integral curve $\gamma_r$ passing through  the   point   $(0,r)$,
with  $r > \frac{1}{2} \ln b$,
at the  initial time. For small values of  $t>0$ the invariant allows to obtain
$x$ in terms of  $y$
\begin{equation} 
ax= - \sqrt{a e^{-y}(e^{r}+ b e^{-r}- e^{y}- b e^{-y})},
\end{equation} 
in such a way that substituting in the second of the equations of the system 
(\ref{ecutrayiso}) it is enough  to do a  quadrature. The   final result
gives the  following expression for  the  curve $\gamma_r$
 \begin{equation}\label{curvagamma} \begin{split}
     \gamma_{r}(t)=& \left(\frac{-(e^{r}- b e^{-r}) S_a(t)
           }{(e^{r}+ b e^{-r})+(e^{r}- b e^{-r}) C_a(t)},
   \ln \frac{1}{2}\left[ (e^{r}+ b e^{-r})+(e^{r}- b e^{-r}) C_a(t) \right] \right).
    \end{split}
  \end{equation} 
From (\ref{curvagamma}) the   flow associated to the system (\ref{ecutrayiso}) is obtained
 supposed $a>0$ and $b>0$
\begin{equation}  \label{flowison}
\begin{split}
   \Phi_{a,b}^t(x,y)=&( \frac{ (
        ax^2e^{y}-e^{y}+b e^{-y})\,S_a(t)+ (2xe^y) \,C_a(t)}{ (ax^2e^{y}+
e^{y}+b e^{-y})+(
        -ax^2e^{y}+e^{y}-b e^{-y})\,C_a(t)+ (2axe^y) \,S_a(t)}, \\[0.3cm]
         & \ln \frac{1}{2}\left[ (ax^2e^{y}+e^{y}+b e^{-y})+(
        -ax^2e^{y}+e^{y}-b e^{-y})\,C_a(t)+ (2axe^y) \,S_a(t)\right] ).
     \end{split}
\end{equation}
It is immediate to
prove that (\ref{flowison}) is also  correct for the remaining values of $a$ and $b$. However,
if the parameters $a$ and $b$ are  positive  the   flow is
defined globally, but this does not happen, in general, for any other value of 
the parameters.

The preceding  study allows to write the   flow
$\Phi_{iN}^t: T_{z, N} \rightarrow  T_{z, N}$
of the vector field $\hat{J}_{iN}$ (\ref{fieldsison}). For its description
it is convenient to introduce the functions
$F^{\omega,z}_{iN}:  T_{z, N} \times \mathbb{R} \rightarrow \mathbb{R}$,
defined by
\begin{equation} \begin{split}
  F^{\omega,z}_{iN}(\alpha, t)=&
        \left[\cosh(z \alpha_N) +\frac{z^2}{2}
        \sum_{s=1}^{N-1} \omega_{sN} \alpha_s^2 e^{z \alpha_N}  \right] \\
      & +    \left[\sinh(z \alpha_N) -\frac{z^2}{2}
        \sum_{s=1}^{N-1} \omega_{sN} \alpha_s^2 e^{z \alpha_N}  \right]     
                C_{\omega_{iN}}(t)+
         \left[z \omega_{iN} \alpha_i e^{z \alpha_N}\right]
                S_{\omega_{iN}}(t).\nonumber
    \end{split}
 \end{equation}
Note that  the   first term can be written in terms of the invariant
$h_{\omega,z}$ as
$$
\cosh(z \alpha_N) +\frac{z^2}{2}
  \sum_{s=1}^{N-1} \omega_{sN} \alpha_s^2 e^{z \alpha_N} =
1+\frac{z^2}{2} h_{\omega,z} (\alpha).
$$
Writing  the  flow action  as
\begin{equation} \label{flujdefison}
   \Phi_{iN}^t(\alpha)= \alpha', 
\end{equation}
we get
\begin{equation}
 \begin{split}
    \alpha'_i= & \frac{
- \left[\sinh(z \alpha_N) -\frac{z^2}{2}
        \sum_{s=1}^{N-1} \omega_{sN} \alpha_s^2 e^{z \alpha_N}  \right]     
                S_{\omega_{iN}}(t)+
         \left[z \alpha_i e^{z \alpha_N}\right]
                C_{\omega_{iN}}(t)}
{ z F^{\omega,z}_{iN}(\alpha,t) }, \\[0.2cm]
    \alpha'_N = & \frac{1}{z} \ln F^{\omega,z}_{iN}(\alpha,t),\\[0.2cm]
   \alpha'_j= & \frac{\alpha_j e^{z \alpha_N}}{F^{\omega,z}_{iN}(\alpha,t)},
\qquad j\not =i, N.\nonumber
 \end{split}\end{equation}
The   limit $z\rightarrow 0$ can be obtained  after considering the 
first order in $z$ of  the 
function $F_{iN}^{\omega, z}$:
$$
 F_{iN}^{\omega, z}(\alpha, t)= 1 + z \left[
  \omega_{iN}  S_{\omega_{iN}}(t)  \alpha_i
  +  C_{\omega_{iN}}(t) \alpha_N \right] + o(z^2) ,
$$
and this result  yields the known linear  flow, 
consisting of `rotations' around the origin of the $iN$ plane,
$$
   \alpha'_i=  C_{\omega_{iN}}(t) \alpha_i  -
                     S_{\omega_{iN}}(t) \alpha_N, \qquad
   \alpha'_N=  \omega_{iN} S_{\omega_{iN}}(t) \alpha_i  +
                     C_{\omega_{iN}}(t) \alpha_N, \qquad
   \alpha'_j=  \alpha_j.
$$


\sect{Example: $U_z(\mathfrak{iso}_{\omega_2, \omega_3}(3))$}
\label{deformacionestandarso2}

In  the  previous section  
$U_z(\mathfrak{iso}_{\omega_2,\omega_3,\ldots, \omega_N}(N))$ has been studied,
now we consider the particular case $N=3$.
The following  discussion clarifies the concepts introduced till now due
to  the  3--dimensional nature  of the group $T_{z 3}$. It is possible
to represent graphically all the  geometric constructions (an enlarged version of this paper 
with some figures can be sent under request to the authors). 

The Hopf algebra 
$U_z(\mathfrak{iso}_{\omega_2, \omega_3}(3))$
is generated by $P_1,\ P_2,\ P_3,\ J_{12}, \ J_{13}$ and $J_{23}$. 
The commutators and the   rest of  structure tensors are obtained after setting the 
corresponding expressions of the
previous   section  for $N=3$. In this case 
$U_z(\mathfrak{iso}_{\omega_2, \omega_3}(3))=
        U_z(\mathfrak{so}_{\omega_2, \omega_3}(3)) \RL F(T_{z,3})$,
where  the group $T_{z,3}$ is characterized by the  composition law
$$
(\alpha'_1, \alpha'_2, \alpha'_3 )
       ( \alpha_1, \alpha_2, \alpha_3 )=
  ( \alpha'_1 + e^{-z \alpha'_3} \alpha_1,
    \alpha'_2 + e^{-z \alpha'_3} \alpha_2, \alpha'_3 + \alpha_3 ).
$$
The translation generators constitute
a  system of global  coordinates over  $T_{z,3}$
$$
  P_1( \alpha_1, \alpha_2, \alpha_3 )= \alpha_1, \quad
     P_2( \alpha_1, \alpha_2, \alpha_3 )= \alpha_2, \quad
      P_3( \alpha_1, \alpha_2, \alpha_3 )= \alpha_3.
$$
Respect to these coordinates
 the  action of  $SO_{\omega_2,\omega_3}(3)$ over  $T_{z,3}$,
induced by the  structure of  
$U_z(\mathfrak{so}_{\omega_2, \omega_3}(3))$--algebra module of  $F(T_{z,3})$, is given 
by the  vector fields
\begin{equation}
\begin{split}
 & \hat{J}_{12}=- P_2 \frac{\partial}{\partial P_1}+\omega_{12} P_1
\frac{\partial}{\partial P_2}, \\[0.3cm]
 & \hat{J}_{13}= -\left[ \frac{1-e^{-2z P_3}}{2z} +
         \frac{z}{2}(\omega_{13} P_1^2- \omega_{23} P_2^2) \right]
                 \frac{\partial}{\partial P_1} -
          z \omega_{13} P_1 P_2       \frac{\partial}{\partial P_2}
           + \omega_{13} P_1      \frac{\partial}{\partial P_3}, \\[0.3cm]
 & \hat{J}_{23}= -z \omega_{23} P_2 P_1  \frac{\partial}{\partial P_1} -
                \left[ \frac{1-e^{-2z P_3}}{2z} +
         \frac{z}{2}(-\omega_{13} P_1^2+ \omega_{23} P_2^2) \right]
                       \frac{\partial}{\partial P_2}
               + \omega_{23} P_2  \frac{\partial}{\partial P_3}.\nonumber
\end{split}
\end{equation}
The  (generalized) distribution  generated by these fields
is integrable since they close the   algebra
$\mathfrak{so}_{\omega_2,\omega_3}(3)$.  The invariant     
$$
        h_{\omega,z}=  \omega_{13} P_1^2 e^{z P_3}+                            
\omega_{23} P_2^2 e^{z P_3}+
     \left[
      \frac{\sinh(\frac{z}{2} P_3)}{\frac{z}{2}}\right]^2
$$
allows us to analyze easily  the  nature of the leaves  of  the  foliation.
The 2--dimensional leaves  correspond to the connected components of  the
sets $h^{-1}_{\omega,z}(t) \subset T_{z,3}$,
being  $t \in \mathbb{R}$ a  regular value of  $h_{\omega,z}$.
For example, when $(\omega_2>0,\ \omega_3>0;\ z>0)$
two strata appear: the   origin point   and the   rest of the space. 
In the   non-deformed case the   study is reduced 
essentially to classify  the  family of  quadrics
$$
\omega_{13} \alpha_1^2 + \omega_{23} \alpha_2^2 +
      \alpha_3^2+ c=0.
$$
When $c\not = 0$ every connected  component constitutes a 2--dimensional orbit of  the 
action, but for $c=0$ zero-dimensional orbits appear.

Last expressions (\ref{flujnodefison}) and (\ref{flujdefison}) allow to describe the
oneparameter flows  associated to the generators. For $\hat{J}_{12}$ it is obtained a  
linear action 
$$
\Phi_{12}^t(\alpha_1, \alpha_2, \alpha_3)=
(C_{\omega_{12}}(t)\alpha_1 -S_{\omega_{12}}(t)\alpha_2,
\omega_{12} S_{\omega_{12}}(t)\alpha_1 +C_{\omega_{12}}(t)\alpha_2 , \alpha_3),
$$
unlike that happens for $\hat{J}_{13}$ and $\hat{J}_{23}$
\begin{equation} \begin{split}
\Phi_{13}^t(\alpha_1, \alpha_2, \alpha_3)= &\left(  \frac{ -\left[\sinh(z \alpha_3)+
   \frac{z^2}{2}(\omega_{13} \alpha_1^2+ \omega_{23} \alpha_2^2)
     e^{z \alpha_3} \right] S_{\omega_{13}}(t) +
      z  \alpha_2 C_{\omega_{13}}(t)}{ z F_{13}(\alpha,t)}\right. ,\\[0.15cm]
        & \qquad \left.\frac{\alpha_2 e^{z \alpha_3}}{F_{13}(\alpha,t)} ,
      \frac{1}{z }  \ln F_{13}(\alpha,t) \right), \\[0.3cm]
     \Phi_{23}^t(\alpha_1, \alpha_2, \alpha_3)= &
       \left(\frac{\alpha_1 e^{z \alpha_3}}{F_{23}(\alpha,t)}\right. , \\[0.15cm]
    &   \qquad \left. \frac{ -\left[\sinh(z \alpha_3)+
   \frac{z^2}{2}(\omega_{13} \alpha_1^2+ \omega_{23} \alpha_2^2)
        e^{z \alpha_3} \right] S_{\omega_{23}}(t) +
            z  \alpha_2 C_{\omega_{23}}(t)}{ z F_{23}(\alpha,t)} ,
       \frac{1}{z }  \ln F_{23}(\alpha,t)\right),\nonumber
   \end{split}
 \end{equation} 
where   
\begin{equation} \begin{split}
     F_{i3}(\alpha,t)= & \left[\cosh(z \alpha_3)+
   \frac{z^2}{2}(\omega_{13} \alpha_1^2+ \omega_{23} \alpha_2^2)
        e^{z \alpha_3} \right]  \\[0.15cm]   
  &\qquad   +   \left[\sinh(z \alpha_3)-
   \frac{z^2}{2}(\omega_{13} \alpha_1^2+ \omega_{23} \alpha_2^2)
        e^{z \alpha_3} \right] C_{\omega_{i3}}(t) +
     z \omega_{i3} \alpha_i  e^{z \alpha_3} S_{\omega_{i3}}(t).\nonumber
   \end{split}
 \end{equation}

The  action of  the oneparameter subgroups gives a new foliation of  the 2--dimensional
leaves presented in the   previous subsection.

The curves that appear in the foliation due, for example, to $\hat{J}_{13}$ may be
interpreted as the  intersection of the surfaces determined by the invariants
$$
\omega_{13} P_1^2 e^{z P_3}+  \omega_{23} P_2^2 e^{z P_3}+
  \left[  \frac{\sin(\frac{z}{2} P_3)}{\frac{z}{2}}\right]^2,
   \qquad   P_2 e^{z P_3}.
$$

Summarizing, all the  qualitative characteristics  relative to  the  deformation 
with respect to the flow of  the fields $\hat{J}_{ij}$, appear in the case $N=3$.

\sect{Concluding remarks}

It is worthy to note that the reinterpretation of the bicrossproduct structure 
${\cal H}= U({\cal K}) \bicross U_z({\cal L})$, in the case that $U_z({\cal L})$ is 
commutative (but noncocommutative) Hopf algebra, as $H= U({\cal K}) \bicross {\rm Fun}({L_z})$
allows to carry the action determining the bicrossproduct to an action of the group $K$ on  $L_z$. 

For the algebras involved in this work, in the deformed case, i.e, $z\neq 0$, the above mentioned
action is local and nonlinear although in the opposite case the action is  global and linear. 

The flows have been obtained studying the case of $\omega_i >0$. An analytical dependence of  the
flow on the parameters $\omega_i$ is observed,  which makes unnecessary to repeat the computations
for the other values of the $\omega_i$'s. This result is very interesting since the structure of
the orbit space of the action of $SO_{\omega_2,\omega_3, \ldots, \omega_N}(N))$ on
$T_{z,N}$ is very complicated, which difficulties to obtain directly the flows for each particular
case. 

In \cite{olmo01a} the flows have been used for the computation of the induced representations for
the $U_z(\mathfrak{iso}_{\omega}(2))$. For higher dimensions the problem of constructing the
induced representations is very cumbersome and it is still an open problem.

The CK family $U_z(\mathfrak{iso}_{\omega_2, \omega_3}(3))$ contains, for instance,  the
$q$--Poincar\'e algebra $(\omega_2 <0, \omega_3>0)$, $(\omega_2 >0, \omega_3<0)$, $(\omega_2 <0,
\omega_3<0)$,   the $q$--Galilei algebra $(\omega_2=0, \omega_3>0)$ and the $q$--Euclidean algebra
$(\omega_2 >0, \omega_3>0)$. For a physical meaning of their generators see Ref.~\cite{olmo94a}.

We finish with the following remarks about the systems (\ref{ecutrayiso}):

\noindent
1) The second order systems associated to  (\ref{ecutrayiso}),
  \begin{equation}
    \begin{split}
        \ddot{x}= - a x \dot{x} -a (1+2x+ax^2) x,  \qquad
        \ddot{y}= -\frac{1}{2} \dot{y}^2- \frac{1}{2}a(1+ be^{-2y}),\nonumber
    \end{split}
  \end{equation}
can be interpreted in both cases as moving objects
over a straight line under  the  action of forces depending on the 
position and  the velocity.

\noindent
2) The   system (\ref{ecutrayiso})
  is associated to the vector field over  $\mathbb{R}^2$
$$
X_{a,b}=  \left[ -\frac{1}{2} a x^2 -\frac{1}{2} +
      \frac{1}{2} b e^{-2y} \right] \frac{\partial}{\partial x}
      + a   x  \frac{\partial}{\partial y}, 
$$
which  admits a  hamiltonian description as we are going to prove.
Obviously, the   pair $(x,y)$ is not a chart of canonical  coordinates since  the  1--form
obtained by contraction  of the vector field and  the  symplectic 2--form  associated to
this chart ($X_{a,b} \rfloor (dx \wedge dy)$)
is not exact.
Hence, let us consider a general symplectic  2--form  
$\omega = \Omega \; dx \wedge dy$,
  with $\Omega$ to be determined.
  Since $h_{a,b}$ is an invariant of the  system it is evident that the Hamiltonian of
the system has to be of  the  form
$h=   f \circ h_{a,b}$,
with $f:\mathbb{R}\rightarrow \mathbb{R}$, which is not univocally determined.
The vector field associated to  $h$ by means  of  the  symplectic structure
is fixed by 
$X_h \rfloor \omega = -dh$. So,
$$
 X_h= - \frac{f'\circ h_{a,b}}{\Omega}
     \; \partial_y h_{a,b} \;
     \frac{\partial }{\partial x} +
      \frac{f'\circ h_{a,b}}{\Omega} \;
        \partial_x h_{a,b} \;
       \frac{\partial }{\partial y}.
$$
Identifying $X_h$ with $X_{a,b}$ two equations are obtained, but only one is 
independent. Hence,
$$
  \Omega=  f'\circ h_{a,b} \frac{\partial_x h_{a,b}}{ax}=
    2 e^{y} f'\circ h_{a,b}.  
$$
The  simple choice  $f(t)=t$ allows us to obtain  the  2--form
$\omega= 2 e^{y}  dx\wedge dy$,
that is independent of  the parameters $a$ and $b$.
With the above election  of  $f$ the Hamiltonian of  $X_{a,b}$
is the invariant $h_{a,b}$.

\section*{Acknowledgments}
This work has been partially supported by DGES of the Ministerio de Educaci\'on y
Cultura de Espa\~na under Project PB98--0360, and the
Junta de  Castilla y Le\'on (Spain).


\end{document}